# Do altmetrics point to the broader impact of research?

# An overview of benefits and disadvantages of altmetrics

Lutz Bornmann

Division for Science and Innovation Studies

Administrative Headquarters of the Max Planck Society

Hofgartenstr. 8,

80539 Munich, Germany.

Email: bornmann@gv.mpg.de

**Abstract**

Today, it is not clear how the impact of research on other areas of society than science should be measured. While peer review and bibliometrics have become standard methods for measuring the impact of research in science, there is not yet an accepted framework within which to measure societal impact. Alternative metrics (called altmetrics to distinguish them from bibliometrics) are considered an interesting option for assessing the societal impact of research, as they offer new ways to measure (public) engagement with research output. Altmetrics is a term to describe web-based metrics for the impact of publications and other scholarly material by using data from social media platforms (e.g. Twitter or Mendeley). This overview of studies explores the potential of altmetrics for measuring societal impact. It deals with the definition and classification of altmetrics. Furthermore, their benefits and disadvantages for measuring impact are discussed.

**Keywords**

Social impact; Broader impact; Altmetrics; Scientometrics

# 1 Introduction

Until a few decades ago, the general assumption in science policy was that a society can benefit most from research that is conducted at a very high level – evaluated according to the standards inherent in science. In recent years, this automatistic approach has found less favour in science policy: policymakers expect science to demonstrate its value to society (Bornmann, 2013). A good example of this trend can be found in a recent book by Bastow, Dunleavy, and Tinkler (2014), which is an attempt to "re-explain the distinctive and yet more subtle ways in which the contemporary social sciences now shape and inform human development" (p. 2). The trend towards audit science is framed in a general change to the science landscape and is frequently described as a development from Mode 1 to Mode 2: While in Mode 1 science was characterized by the academic interests of a scientific community, Mode 2 is more concerned with the collaboration between science and other areas of society and with research that is relevant to a particular application in society (Gibbons et al., 1994).

It is not clear how the impact of research on other areas of society should be measured – unlike the impact which research has on itself. While peer review and bibliometrics have become standard methods for measuring the impact of research on other research, there is not yet an accepted framework within which to measure societal impact. Nowadays, the case study approach to societal impact is favoured; however, this approach does not meet all the requirements generally associated with a societal impact framework. According to Frank and Nason (2009), the best method of measuring societal impact (in health research) should be "feasible, not too labour intensive, and economically viable. It should be as accurate and responsive as possible within a reasonable evaluation budget that should represent a small percentage of the money invested in the research being assessed" (p. 531). There is a need for indicators which can reliably and validly measure the impact of research on certain parts of society, with the primary aim of creating productive interaction and successful communication between research and societal stakeholders. "Scientists must be able to explain what they do to a broader public to garner political support and funding for endeavours whose outcomes are unclear at best and dangerous at worst, a difficulty which is magnified by the complexity of scientific issues" (Puschmann, 2014, p. 91).

## 2      What are altmetrics?

Alternative metrics (called altmetrics to distinguish them from bibliometrics, Gunn, 2013) are considered an interesting option for assessing the societal impact of research, as they offer new ways to measure (public) engagement with research output (Piwowar, 2013). "Altmetrics … is a term to describe web-based metrics for the impact of scholarly material, with an emphasis on social media outlets as sources of data" (Shema, Bar-Ilan, & Thelwall, 2014).[1] In 'article-level metrics' (ALMs, Fenner, 2013b), views, downloads, clicks, notes, saves, tweets, shares, likes, recommends, tags, posts, trackbacks, discussions, bookmarks, and comments are counted, rather than just citations of a paper in a database such as Scopus (Elsevier), or by a publisher such as the Public Library of Science (PLOS, Fenner, 2013b) (Liu, Xu, Wu, Chen, & Guo, 2013; Zahedi, Costas, & Wouters, 2014). Adie and Roe (2013) call these individual events (tweets or shares, for example) 'mentions' if they link to papers (and 'posts' if they do not). <u>Every form</u> of ALM involves log data which measures individual mentions over a certain period of time (Haustein, 2014). "Today, for every single use of an electronic resource, the system can record which resource was used, who used it, where that person was, when it was used, what type of request was issued, what type of record it was, and from where the article was used" (Kurtz & Bollen, 2010, p. 4). The more or less frequent

---

[1] Rousseau and Ye (2013) have proposed "influmetrics" as a new name for this new form of metrics. Cronin (2013) thinks that "complementary metrics" is more appropriate than "alternative metrics". "Influmetrics" has the advantage against "complementary metrics" and "altmetrics" that it does not provoke the question "complementary or alternative to what?"

"use" of research output can either be seen as the direct impact of research or as evidence of "real" impact (Neylon, Willmers, & King, 2014).

The importance of this alternative form of metrics is indicated by one of the biggest multidisciplinary database providers, Elsevier, not only entering into partnership with Altmetric, a start-up tracking and analysing the online activity around scholarly literature, but also buying Mendeley, which combines a citation manager with a scholarly social network (Roemer & Borchardt, 2013). Furthermore, according to Chamberlain (2013) and Piwowar and Priem (2013), scholars are already including altmetrics in publication lists in their CVs (in addition to citation impact measurements), conferences on the subject are being arranged (such as altmetrics.org/altmetrics14) and organizations (such as ImpactStory and Altmetric) founded to collect and provide altmetrics (Fenner, 2013a). Against the background of this development Bornmann (2014) and Taylor (2013a) are talking about a revolution in scientometrics, Lin and Fenner (2013) about a new paradigm of research assessment and Kurtz and Bollen (2010) about a renaissance in bibliometrics with, notably, a new definition of the expression "impact of science". According to Galloway, Pease, and Rauh (2013) "altmetrics is a fast-moving and dynamic area".

However, the use of alternative metrics to evaluate research is not new. It has a long tradition in scientometrics with the analysis of acknowledgements, patents, mentorships, news articles, and usage in syllabi (Priem, 2014). The use of the Internet for alternative metrics began with "webometrics" (or "cybermetrics") whereby the number of times a paper was mentioned on the web was counted (Roemer & Borchardt, 2012). These mentions were called "web citations" (Shema, Bar-Ilan, & Thelwall, in press). Several studies have investigated the relationship between web citations and traditional text citations finding moderate correlations in most cases (see e.g. Kousha & Thelwall, 2007; Vaughan & Shaw, 2005; Vaughan & Shaw, 2008).

Later on, server download data for papers was analysed in order to measure (scientists') interest in papers (Gunn, 2013). The development of the Internet into social web – as a new social media platform – also led to new ways to measure impact. The social web is characterized by many applications which promote participation, interconnections, social interaction and user-generated content (Greenhow & Gleason, 2014; Weller & Peters, 2012). The user of content in the social web not only consumes but also provides it and comments on it (King et al., 2013).

In scientometrics, the focus has been moving from web citation analysis (and the analysis of download data) towards social media usage analysis (Li, Thelwall, & Giustini, 2012), known currently as "altmetrics". In recent years, the use of the following seven platforms in the social web as alternative metrics is of primary interest: "bookmarking, reference managers, recommendation services, comments on articles, microblogging, Wikipedia, and blogging" (Priem & Hemminger, 2010). These platforms provide an insight into the research process, as the data, analyses, and results can be exchanged, stored and discussed (Fausto et al., 2012). However, alternative metrics are (still) greatly in flux, with new tools being considered as data sources and established tools losing their appeal as data sources (Darling, Shiffman, Côté, & Drew, 2013). Over recent years, however, a number of tools have proved particularly suitable for alternative measurement: according to Fenner (2013b), for example, since June 2012 93% of PLOS Biology papers have been mentioned on Twitter.

## 3  How can altmetrics be classified?

As there are now a number of social media tools which can be used as a source for altmetrics (see e.g. the long list of Claussen et al., 2013, p. 360), they have been classified by authors into various areas. This classification not only provides an overview of the different metrics, but also indicates the type of application for which each metric is suitable. Two possible forms of classification used by ImpactStory and PLOS are described in Table 1. As the classified metrics are generally ALMs and not only altmetrics, the category "cited" – as in "cited by scientists" – is also listed.

Table 1. The ALM classification of ImpactStory and PLOS (Lin & Fenner, 2013)

| Area | Scholars | Public |
| --- | --- | --- |
| **ImpactStory** | | |
| Viewed | PDF downloads | HTML downloads |
| Saved | CiteULike, Mendeley | Delicious |
| Discussed | Science blogs, journal comments | Blogs, Twitter, Facebook |
| Recommended | Citations by editorials | Press article |
| Cited | Citations, full-text mentions | Wikipedia mentions |
| **PLOS** | No distinction is made between scholars and public | |
| Viewed | HTML/ PDF (PLOS or PubMed Central), XML (PLOS) | |
| Saved | CiteULike, Mendeley | |
| Discussed | NatureBlogs, ScienceSeeker, ResearchBlogging, PLOS Comments, Wikipedia, Twitter, Facebook | |
| Recommended | F1000Prime | |
| Cited | CrossRef, PubMed Central, Web of Science, Scopus | |

Both classifications cover viewing, storing, discussing, recommending and citing papers (or other products of research) and therefore mirror the whole process of user engagement, from the first look at a paper to its citation in (scholarly) literature and thus relate to the various dimensions of research impact (Neylon & Wu, 2009). In this process of user engagement, it is expected that the number of counts per paper falls (from viewing to storing, discussing, recommending and citing) and the significance of individual mentions increases (with citations having the largest significance) (Kurtz & Bollen, 2010). According to Lin and Fenner (2013), only one person in 70 cites a paper that they have downloaded from PLOS in their own paper.

With the exception of one type of ALM in Table 1, all the ALMs are assigned in the same way to the two classifications: While Wikipedia is included under "cited" by ImpactStory, PLOS has it in the "discussed" category. The two classifications differ significantly in that ImpactStory distinguishes between impact on scholars (scientific impact) and the public (societal impact). However, this distinction is sometimes quite artificial: PDFs are not only downloaded by scholars and HTML versions not only by the public. There are specific advantages and disadvantages to each of the metrics listed in the table in their measurement of impact. For example, comments can allow valuable and rapid feedback to a paper; however, they are not given frequently enough on individual papers to be used validly as a metric (Neylon & Wu, 2009).

Plum Analytics – a supplier of impact metrics similar to ImpactStory – also uses a classification similar to that in Table 1, with usage, captures, mentions, social media, and citations. Further classifications for altmetrics are as follows: Haustein and Peters (2012) distinguish between Web data (e.g. tweets, bookmarks and blog posts) and Web tools (e.g. social bookmarking systems and reference managers). While the various Web tools can be categorized as sharing services (e.g. YouTube and Flickr) and social bookmarking services (e.g. Delicious) (Haustein, 2014), the social networks (such as Facebook and Twitter) can be divided into informal and formal networks (Rodgers & Barbrow, 2013). Gunn (2013) designates some altmetrics as content-rich (e.g. blog posts or Wikipedia links) and others as plentiful or content-poor (e.g. tweets or Facebook's "like").

# 4 What benefits do altmetrics offer?

The following list of the benefits of altmetrics is based on a categorisation of the benefits named in the literature by Wouters and Costas (2012). In an overview of new forms of impact measurements, these authors identified four benefits that altmetrics has compared to traditional metrics: (1) Broadness: altmetrics measure impact beyond science. (2) Diversity: altmetrics can measure the impact of scholarly products other than papers. (3) Speed: altmetrics permit impact to be measured shortly after the publication of a paper (or the completion of other products). (4) Openness: as a rule, it is easy to obtain altmetric data.

## 4.1 Broadness

Most comments on the benefits of altmetrics relate their potential for measuring the broader impact of research, that is, beyond science (Priem, Parra, Piwowar, & Waagmeester, 2011; Priem, Piwowar, & Hemminger, 2012; Weller, Dröge, & Puschmann, 2011). It is hoped that altmetrics can deliver more transparent descriptions of the interest, usage and reach of scholarly products (Fausto, et al., 2012; Taylor, 2013a) and also more diverse and nuanced forms of impact analyses than traditional metrics, such as bibliometrics, permit (Waltman & Costas, 2014). Statements such as those by Mohammadi, Thelwall, Haustein, and Larivière (2014) are pertinent: "A noticeable percentage of Clinical Medicine papers were read by people who are apparently not academics and this is an important issue because some articles could be useful in clinical practice even if they are not cited in the literature." As citations only relate to the assessment by scientific authors of the research conducted by fellow researchers, altmetrics offer access to the opinions of a wider audience, such as professionals, undergraduates, government and – as a whole – the interested general public (Adie, 2014; Hammarfelt, 2014).

The potential to measure the flow of research into society complies with the wishes of politicians, research organisations and funders, such as the broader impact criteria required by the US National Science Foundation (Chamberlain, 2013). Konkiel and Scherer (2013) propose altmetrics as a supplementary indicator of impact with which to justify budget increases and recruiting faculty for university trustees and state legislatures. Some authors, such as Bik and Goldstein (2013), ascribe the potential to measure the "true" or "full" impact of research to altmetrics. However, this is an exception in the literature and undoubtedly overstates the possibilities they offer. "Hidden impact", the term used by Taylor (2013b), is more fitting, with its implication that altmetrics could reveal impact which traditional indicators have hitherto been unable to reveal. According to Fenner (2013b), altmetrics allow the impact of research to be measured in more practical fields, and papers of general interest to be highlighted better than with citations. This covers forms of impact, such as policy change, and effects on clinical practice, technical applications, education, and health policies (Haustein, 2014; Haustein et al., 2014; Neylon, et al., 2014).

## 4.2 Diversity

Altmetrics are not only more diverse in kinds of data (see above) and accordingly numbers of data sources (whereas for traditional citations only the cited references in journals serve as data source), but also allow for evaluation of a greater diversity of products, i.e., not just publications. Research funders, such as the US National Science Foundation, expect meanwhile not only publications but also other products to be given as the outcome of research in proposals. This new requirement should be understood as an indication that not only publications but also other forms of scholarly products play an important part in research evaluation now (Piwowar, 2013; Rousseau & Ye, 2013). With reference to evaluating these other products, the proviso imposed by the US National Science Foundation that only citable and accessible products can "count" is crucial. These products might be datasets, software, copyrights, algorithms, grey literature, and slides (Zahedi, et al., 2014). Altmetrics now offer the opportunity to determine the impact of these products both in science – they are usually under-represented in the citation record (Priem, 2014) – and beyond science (Galloway, et al., 2013). As well as measuring the impact of products, altmetrics can also be used to track a variety of scholarly activities such as teaching and service activities (Rodgers & Barbrow, 2013). For example, the impact of course packs and reading lists or attendance at online open courses (MOOCs) can be measured (Taylor, 2013a).

**4.3     Speed**

One of the biggest disadvantages of citation counts in measuring impact is that a reliable and valid measurement can only be provided several years after publication (Wang, 2013). Altmetrics, on the other hand, permit the impact of a paper (or other products) to be measured just a few days or weeks after it has appeared (Haustein, Peters, Bar-Ilan, et al., 2014; Mohammadi & Thelwall, 2014). For example, the results of Maflahi and Thelwall (in press) suggest that papers tend to attract more Mendeley readers than citations initially, but that the situation reverses after several years. Relatively soon after publication, a paper is read, bookmarked, saved, annotated and discussed within academic circles and by the public (Priem, 2014; Rodgers & Barbrow, 2013). The prompt tweeting or blogging of research results can even assist scientists to secure priority for the results before they are submitted to a journal, on the basis of a preprint (Darling, et al., 2013). Many social web tools offer real-time access to structured altmetric data via application programming interfaces (APIs) (Priem & Hemminger, 2010), with which the impact of a paper can be tracked at any time after publication. This real-time access can be used by scientists and others to track online activities on certain research topics of interest in order to obtain references to important studies which have just been published (Priem, Taraborelli, Groth, & Neylon, 2010).

**4.4     Openness**

A major problem with the societal impact analyses undertaken up to now has been the availability of data. While citation counts for impact measurements in science are available in multi-disciplinary databases (such as the Web of Science, Thomson Reuters, and Scopus), there has been no such easily accessible broad-based data for measuring societal impact. For this reason, the case study approach was favoured for measuring societal impact, whereby it was only measured case-specific and not standardized (Bornmann, 2012, 2013). Altmetrics represent an interesting option for measuring societal impact instead of a case study. In particular, free access to this data through Web APIs, which allow immediate feedback about a large publication set (Galloway, et al., 2013) means that data collection is less problematic (Thelwall, Haustein, Lariviere, & Sugimoto, 2013). Furthermore, altmetric data is today based on platforms with clearly defined boundaries and data types, as is the case with Twitter or Mendeley (Priem, 2014), which facilitates the analysis of data and the interpretation of results.

## 5 What are the disadvantages of altmetrics?

It goes without saying that altmetrics have disadvantages as well as advantages. They share this characteristic with traditional metrics. Not everything that is cited has been read, and the relevant publications are not always cited in the correct place in a manuscript (Haustein, 2014). Furthermore, there are numerous different reasons why scientists cite a publication – and they are not always related to intellectual influence (Bornmann & Daniel, 2008). For Priem (2014), a lack of theory, ease of gaming, and possible biases are three limitations of altmetrics. The following discussion generalizes and adds to this list.

### 5.1 Commercialisation

As commercial providers, many services in the social media (such as Twitter and Facebook), have a large stake in as many people as possible communicating as often as possible via their portals. For example, e-mails constantly draw the attention of the users of these portals to other potentially interesting users and content. A lack of communication or an unwillingness to communicate could result in the portal's failure to thrive commercially. So far, there have been no empirical studies to investigate how much bias this promotion of communication creates for altmetrics. This kind of commercialisation plays next to no part in traditional metrics, such as bibliometrics. Scientists are not encouraged to cite as much as possible. Although publication and citation figures are made available in the Web of Science and Scopus, neither Thomson Reuters nor Elsevier pursues strategies to increase the number of publishing or citing authors.

**5.2 Data quality**

There are a number of different aspects concerning data quality which could lead to a limitation of altmetrics:

1. Bias: As not everyone (in a city, a country, etc.) uses social media platforms, a measurement of impact always relates to a specific sample of people who have mentioned a paper more or less frequently. It is assumed that this sample has a systematic bias towards younger or more fad-embracing people (Priem, 2014) or towards those with a professional interest in research (Neylon, et al., 2014). As there are no accurate user statistics or sample descriptions for individual social media platforms, this bias cannot be quantified.

2. Target: Altmetric counts are frequently made available as counts of all relevant mentions on a platform. However, more information about user groups who have had to do with a scientific paper is essential for a valid measurement of societal impact: Has impact been measured in government documents or on social media comment sites (Liu & Adie, 2013)? This more specific description of the impact achieved is usually lacking nowadays.

3. Multiple versions: Publications often exist in different versions (for example as pre-prints on arXiv.org and post-prints from a publisher). As a result, using altmetrics to measure their impact results in ambiguity and redundancy (Liu & Adie, 2013).

4. <u>Different meanings</u>: Citations might be simple mentions or extensive discussions of a cited paper. Meaning is similarly expanded when applied to social media conversations. These can be very technical and detailed, or also consist merely of a simple mention (Neylon, et al., 2014; Taylor, 2013b). It would be desirable to have the different meanings taken into account in the measurement.

5. <u>Measurement standards:</u> Every scientist knows what is being measured with a citation count: the number of times a paper is listed in the references of subsequently published documents. In altmetrics it is often not clear what is being measured – even if the source for the metric is the same. The respective numbers can refer to different forms of engagement, as in the example given by Liu and Adie (2013): "Quantification of the mentions of scholarly articles on Facebook could take into account either all or just public wall posts, and these posts might be further parsed into the number of wall posts with an article mention or the number of 'likes' and comments on that wall post. Each number emphasizes something different and thus paints a slightly different picture of engagement with an article on Facebook" (p. 32).

6. <u>Mention standards:</u> There are precise rules governing when, where and in which form papers are cited in a document (American Psychological Association, 2010), although not all scientists comply with them. There are no similar rules applying to the various social media platforms (Taylor, 2013b). This means that many links to the research under discussion are included in the text in different ways or not at all (Neylon, et al., 2014). This makes it significantly more difficult to count mentions of papers on these platforms. Providers of altmetric data, such as Altmetric, try to solve the problem with text-mining mechanisms (Liu & Adie, 2013).

7. <u>Normalization:</u> Citations are normalized to allow cross-field and cross-time comparisons of the impact of papers (Bornmann, Leydesdorff, & Mutz, 2013; Vinkler, 2010). As higher altmetric scores can be expected from newer papers and papers on certain topics (such as evolution or climate change) than for older papers and papers on other topics, altmetric data should also be normalized (Holmberg & Thelwall, 2014; Taylor, 2013a; Thelwall, et al., 2013). Only normalized scores allow the (societal) impact of papers on different topics and from different periods to be compared (Torres-Salinas, Cabezas-Clavijo, & Jimenez-Contreras, 2013). Up to now, it has not been common to normalize altmetrics, but ImpactStory already does so on the basis of percentiles (Chamberlain, 2013; Roemer & Borchardt, 2013).

8. <u>Replication:</u> Citation numbers from the Web of Science and Scopus can be replicated as a rule – if one takes into account that the numbers rise over time. The replication of altmetric data is difficult, as data providers change, become quickly obsolete or changes are made to the service they offer (Haustein, Peters, Sugimoto, Thelwall, & Larivière, 2014; Thelwall, et al., 2013). Particularly when altmetrics are to be used to evaluate research, it is important that the results can be replicated. Results of research evaluation often lead to critical discussions.

**5.3    Missing Evidence**

The lack of evidence of altmetrics relate to the scarcity of sophisticated empirical studies on altmetrics. Meaningful results can only be obtained with application of advanced empirical methods. According to Haustein, Peters, Sugimoto, et al. (2014) "large-scale studies of altmetrics are rare, and systematic evidence about the reliability, validity, and context of these metrics is lacking" (p. 657). Many studies which have been conducted on altmetrics up to now employ inadequate methods. Samples are collected randomly or according to the 'snowball' principle; correlation coefficients are not interpreted in accordance with established guidelines; many statistical tests on the same dataset are conducted without a correction to the alpha level (the Bonferroni correction, for example) and the statistical significance is used as the (sole) criterion for the importance of results (Bornmann & Williams, 2013). In a content analysis, of a blog for example, the reliability of the assigned categories is not measured as a rule. It is only possible to determine inter-coder reliability with a comparison of the categorizations by two coders.

**5.4     Manipulation**

It is much easier to manipulate altmetrics than bibliometrics (Rousseau & Ye, 2013; Thelwall, et al., 2013). Regarding traditional metrics, there are reports that journals try to increase their impact with several citations in editorials or that Google Scholar can create citations with false papers (Delgado López-Cózar, Robinson-García, & Torres-Salinas, 2014). However, there are many more and different opportunities for manipulation with altmetrics that are much easier to carry out. "In particular, since social websites tend to have no quality control and no formal process to link users to offline identities it would be easy to systematically generate high altmetric scores for any given researcher or set of articles" (Thelwall, 2014, p. 4). For example, Twitter mentions can be generated through fake accounts and "robot tweeting" (Darling, et al., 2013; Liu & Adie, 2013).

A possible measure to counter manipulation of altmetrics is the cross-calibration of data from different sources in order to reveal suspicious patterns in a source (Priem & Hemminger, 2010).

# 6   Discussion

The significance ascribed to alternative metrics for the evaluation of research fluctuates. While Haustein, Peters, Sugimoto, et al. (2014) already view these metrics as a fixed part of research evaluation, Cronin (2013) is sceptical: "Neither Twitter mentions nor Facebook 'likes' are, for now at any rate, accepted currencies in the academic marketplace; you are not going to get promoted for having been liked a lot, though it may well boost your ego. A robust h-index, on the other hand, could work wonders for your career" (p. 1523). Even if there is no conclusive evidence of the significance of altmetrics for research evaluation, it is clear that research on and the use of altmetrics is becoming more and more popular and the (critical) discussions about possible application scenarios are increasing (Peters, Beutelspacher, Maghferat, & Terliesner, 2012). One gains the impression thereby that altmetrics is not a short-lived object of study in the information sciences, but is establishing itself as a new subfield (Priem, et al., 2010). The growing significance of altmetrics (in the information sciences) is also noticeable from the number of overviews on research into the subject which look at the area from different perspectives (Bar-Ilan, Shema, & Thelwall, 2014; Galloway, et al., 2013; Haustein, 2014; Priem, 2014; Rodgers & Barbrow, 2013; Torres-Salinas, et al., 2013; Wouters & Costas, 2012).

In a survey of bibliometricians, around 86% of those surveyed said that they thought altmetrics had some potential for author or article evaluation. Compared with paper downloads or views, for which 72% see some potential, the potential of typical altmetric platforms, such as blogs or bookmarks on reference managers, is given at round 35%, which, however, is significantly lower (Haustein, Peters, Bar-Ilan, et al., 2014). This study has also concluded that altmetrics offer great potential (and many expectations are associated with altmetrics); however, there are a number of problems which must be solved before it is used to evaluate research. According to Zahedi, et al. (2014) "the study of altmetrics is in its early stage" and Taylor (2013b) says that "little is known about the intentional, motivational or experiential motives of the users" (p. 19). Many of the studies published so far have merely calculated correlations between citations and altmetrics (see e.g. Eysenbach, 2011). However, little knowledge is gained from these studies. The correlation on a medium level found by most studies is hardly meaningful and can be interpreted fairly loosely. As there is no interest in replacing traditional bibliometrics with altmetrics, research should not concentrate on the connectedness, but on specific differences between the two metrics (Darling, et al., 2013): in how far can altmetrics – unlike the traditional metrics – measure the broader impact of research?

The literature gives an abundance of issues with which research into altmetrics should concern itself in future. The white paper by the National Information Standards Organization (2014) in particular offers a number of recommendations as to which research is important in this area. Lin and Fenner (2013) emphasize, for example, the development of sophisticated technologies to analyse the demographics of research users more accurately. There is great interest in finding out whether users are scholars or non-scholars, and how they are distributed geographically and what stages they have reached in their careers. Where the broader impact of research is concerned, it is much more important to learn who has used an actual research product and why, than to simply know "how many" people have in total.

According to Priem, et al. (2010), in order to be able to answer the question of whether altmetrics measure impact ("or just empty buzz"), it should be compared with expert evaluations (Sud & Thelwall, 2014): Do altmetric counts correlate with the evaluations by experts of the societal impact of a paper? A good source of data for studying this correlation is F1000 Prime (Mohammadi & Thelwall, 2013). What is known as the Faculty of 1000 (F1000) peer review system is accordingly not an ex-ante assessment of manuscripts provided for publication in a journal, but an ex-post assessment of papers which have already been published in journals. The Faculty members also attach tags to the papers indicating their relevance for science (e.g. "new finding" or "good for teaching"). The ex-post assessments and tags can be used for the investigation of altmetrics.

In scientometrics, procedures should be developed to detect and repair gaming in altmetrics. Furthermore, attention should be given to the problem of representativeness of altmetric data (Haustein, 2014): For example, if one would like to measure the impact of a research product on politics, one needs to know how strongly politics is represented on a certain social media platform.

Fundamentally, it should be ensured that when altmetrics are used in research evaluation, they are in an informed peer review process, exactly like the traditional metrics. Results based on altmetrics must therefore not lead directly to decisions about research funding, but should be used to help experts to make decisions in a peer review process (Bornmann, 2011; Rousseau & Ye, 2013). The traditional and alternative metrics should complement (and not replace) each other in an informed peer review process.